\documentclass[natbib]{svjour3}
\pdfoutput=1
\usepackage{etex}

\usepackage[T1]{fontenc} 
\usepackage[utf8]{inputenc} 
\usepackage[table,xcdraw]{xcolor}
\usepackage{graphicx}

\usepackage{balance}
\usepackage{paralist}
\usepackage{pdflscape}
\usepackage[]{pdfcomment}
\newcommand{\TODO}[1]{\textcolor{red}{#1}\pdfcomment[color=yellow,open=false]{#1}}\newcommand\todo\TODO

\newcommand{\mycode}[1]{{\small \texttt{#1}}\xspace}

\newcommand{\jgenprog}{jGenProg\xspace}
\newcommand{\jkali}{jKali\xspace}

\newcommand{\nbAllBugs}{224\xspace}

\newcommand{\nbFixedBugs}{47\xspace} 
\newcommand{\nbFixedBugsByKali}{22\xspace}
\newcommand{\nbFixedBugsByNopol}{35\xspace}
\newcommand{\nbFixedBugsByGenprog}{27\xspace}
\newcommand{\nbFixedBugsByAll}{12\xspace}
\newcommand{\nbDifferentPatches}{84\xspace}
\newcommand{\nbAnalyzedPatchCorrect}{11\xspace}
\newcommand{\nbAnalyzedBugCorrect}{9\xspace}
\newcommand{\nbAnalyzedGenprogCorrect}{5\xspace}
\newcommand{\nbAnalyzedIncorrect}{61\xspace}
\newcommand{\nbAnalyzedUnknown}{12\xspace}
\newcommand{\nbAnalyzedReaEasy}{61\xspace}
\newcommand{\nbAnalyzedDiffHardExpert}{21\xspace}
\newcommand{\experimenttimeindays}{17.6\xspace}
\newcommand{\averageExecutionTimeWithPatch}{14.8\xspace}
\newcommand{\blind}[1]{}

\usepackage{comment}

\usepackage[]{hyperref}
\usepackage{xspace}
\def\ourif{{\sc if}\xspace}

\usepackage{listings}
\usepackage{multirow}
\usepackage{amsmath}

\title{
Automatic Repair of Real Bugs in Java: \\ A Large-Scale Experiment on the Defects4J Dataset
}
\titlerunning{Automatic Repair of Real Bugs in Java}

\author{Matias~Martinez \and Thomas~Durieux \and Romain~Sommerard \and Jifeng~Xuan \and Martin~Monperrus}
\authorrunning{Martinez, Durieux, Sommerard, Xuan, Monperrus }

\institute{M.~Martinez
\at University of Lugano,  Via Giuseppe Buffi 13, 6900 Lugano, Switzerland\\Tel.:  +41 58 666 40 00
\\\email{matias.sebastian.martinez@usi.ch}
\and J.~Xuan
\at State Key Lab of Software Engineering, Wuhan University, 299 Bayi Road, 430072 Wuhan, China \\Tel.: +86 27 6877 6139\\\email{jxuan@whu.edu.cn}
\and
T.~Durieux \and R.Sommerard \and M.~Monperrus
              \at INRIA \& University Lille 1,  40 Avenue du Halley, 59650 Villeneuve-d'Ascq, France \\Tel.:  +33 03 59 57 78 00
\\\email{thomas.durieux@inria.fr}
\\\email{romain.sommerard@etudiant.univ-lille1.fr}
\\\email{martin.monperrus@univ-lille1.fr}
}

\begin{document}
\maketitle

\begin{abstract} 
Defects4J is a large, peer-reviewed, structured dataset of real-world Java bugs. Each bug in Defects4J comes with a test suite and at least one failing test case that triggers the bug.
In this paper, we report on an experiment to explore the effectiveness of automatic test-suite based repair on Defects4J.
The result of our experiment shows that the considered state-of-the-art repair methods can generate patches for 
\nbFixedBugs out of \nbAllBugs bugs.
However, those patches are only test-suite adequate, which means that they pass the test suite and may potentially be incorrect beyond the test-suite satisfaction correctness criterion.
We have manually analyzed \nbDifferentPatches different patches to assess their real correctness.
In total, \nbAnalyzedBugCorrect real Java bugs can be correctly repaired with test-suite based repair.
This analysis shows that test-suite based repair suffers from under-specified bugs, for which trivial or incorrect patches still pass the test suite.
With respect to practical applicability, it takes on average \averageExecutionTimeWithPatch minutes to find a patch.
The experiment was done on a scientific grid, totaling \experimenttimeindays days of computation time. All the repair systems and experimental results are publicly available on Github in order to facilitate future research on automatic repair.
\end{abstract}


\section{Introduction}

Automatic software repair is the process of automatically fixing bugs.
Test-suite based repair, notably introduced by GenProg (\cite{le2012genprog}), consists in synthesizing a patch that passes a given test suite, which initially has at least one failing test case. 
In this paper, we say that this patch is ``test-suite adequate''.
In this recent research field, few empirical evaluations have been made to evaluate the practical ability of current techniques to repair real bugs.

For bugs in the Java programming language, one of the largest evaluations is by \cite{Kim2013}, but as discussed in \cite{monperrus2014critical}, this evaluation suffers from a number of biases and cannot be considered as a definitive answer to the repairability of Java bugs.
For bugs in the C programming language, according to bibliometrics, the most visible one is by \cite{le2012systematic}. They reported on an experiment where they ran the GenProg repair system on 105 bugs in C code.
However, \cite{qi2015efficient} have shown at ISSTA 2015 that this evaluation suffers from a number of important issues, and call for more research on systematic evaluations of test-suite based repair. 

\emph{Our motivation is to conduct a novel empirical evaluation in the realm of Java bugs, in order to better understand the fundamental strengths and weaknesses of current repair algorithms.}

In this paper, we present the results of an experiment consisting of running repair systems for Java on the bugs of Defects4J. 
Defects4J is a dataset (\cite{JustJE2014}) by the University of Washington consisting of 357 real-world Java bugs. It has been peer-reviewed, is publicly available, and is structured in a way that eases systematic experiments. Each bug in Defects4J comes with a test suite including both passing and failing test cases.

For selecting repair systems, our inclusion criteria is that 1) it is publicly available and 2) it runs on modern Java versions and large software applications. This results in \jgenprog, an implementation of GenProg (\cite{le2012genprog} for Java; \jkali, an implementation of Kali (\cite{qi2015efficient}) for Java; and Nopol (\cite{demarco2014automatic,nopoljournal}). 

Our experiment aims to answer to the following Research Questions (RQs):

\textbf{RQ1}. \textit{Can the  considered repair systems synthesize patches for the bugs of the Defects4J dataset?}
Beyond repairing toy programs or seeded bugs, for this research field to have an impact, one needs to know whether the current repair algorithms and their implementations work on real bugs of large scale applications.  

\textbf{RQ2}. \textit{In test-suite based repair, are the generated patches correct, beyond passing the test suite?}
By ``correct'', we mean that the patch is meaningful, really repairing the bug, and is not a partial solution that only works for the input data encoded in the test cases.
Indeed, a key concern behind test-suite based repair is whether test suites are good enough to drive the generation of correct patches, where correct means acceptable. Since the inception of the field, this question has been raised many times and is still a hot question: \cite{qi2015efficient}'s recent results show that most of GenProg's patches on the now classical GenProg benchmark of 105 bugs are incorrect. We will answer RQ2 with a manual analysis of patches synthesized for Defects4J. 

\textbf{RQ3}. \textit{Which bugs in Defects4j are under-specified?}
A bug is said to be under-specified if there exists a trivial patch that simply removes functionality.
For those bugs, current repair approaches fail to synthesize a correct patch due to the lack of test cases. Those bugs are the most challenging bugs: to automatically repair them, one needs to reason on the expected functionality below what is encoded in the test suite, to take into account a source of information other than the test suite execution. One outcome of our experiment is to identify those challenging bugs.

\textbf{RQ4}. \textit{How long is the execution of each repair system?}
The answer to this question also contributes to assess the practical applicability of automatic repair in the field.

Our experiment considers \nbAllBugs bugs that are spread over 231K lines of code and 12K test cases in total. We ran the experiment for over \experimenttimeindays days of computational time on Grid'5000 (\cite{grid5000}), a large-scale grid for scientific experiments. 

Our contributions are as follows:

\begin{itemize}

\item \textbf{Answer to RQ1}. The Defects4J dataset contains bugs that can be automatically repaired with at least one of the considered systems. \jgenprog, \jkali, and Nopol together synthesize test-suite adequate patches for \nbFixedBugs out of \nbAllBugs bugs with \nbDifferentPatches different patches. Some bugs are repaired by all three considered repair approaches (\nbFixedBugsByAll/\nbFixedBugs). \textbf{These results validate that automatic repair works on real bugs in large Java programs\footnote{The dataset and the repair system in \cite{Kim2013} are not publicly available.}. They can be viewed as a baseline for future usage of Defects4J in automatic repair research.}

\item \textbf{Answer to RQ2}. Our manual analysis of all \nbDifferentPatches generated patches shows that \nbAnalyzedPatchCorrect/\nbDifferentPatches are correct, \nbAnalyzedIncorrect/\nbDifferentPatches are incorrect, and \nbAnalyzedUnknown/\nbDifferentPatches require a domain expertise, which we do not have. The incorrect patches tend to overfit the test cases. This is a novel piece of empirical evidence on real bugs that either the current test suites are too weak or the current automatic repair techniques are too dumb. \textbf{These results strengthen the findings in \cite{qi2015efficient}, on another benchmark, on another programming language. They show that the overfitting problem uncovered on the Genprog's benchmark by \cite{qi2015efficient} is not specific to the benchmark and its weakness, but is more fundamental.}

\item \textbf{Answer to RQ3}. Defects4J contains very weakly specified bugs. Correctly repairing those bugs can be considered as the next milestone for the field.
\textbf{This result calls for radically new automatic repair approach that reasons beyond the test suite execution, using other sources of information, and for test suite generation techniques tailored for repair.} 

\item \textbf{Answers to RQ4}. The process of searching for a patch is a matter of minutes for a single bug. 
\textbf{This is an encouraging piece of evidence that this research will have an impact for practitioners.} 

\end{itemize}

For the sake of open science and reproducible research, our code and experimental data are publicly available on Github (\cite{githubresults,astorcode,nopolcode}).

The remainder of this paper is organized as follows. Section \ref{sect:background} provides the background of test-suite based repair and the dataset. Section \ref{sect:protocol} presents our experimental protocol. Section \ref{sect:result} details answers to our research questions. Section \ref{sect:case-study} studies three generated patches in details. Section \ref{sect:discussion} discusses our results and Section \ref{sect:related} presents the related work. Section \ref{sect:conclusion} concludes this paper and proposes future work.

\begin{figure}[!t]
\centering
\includegraphics[width=0.48\textwidth]{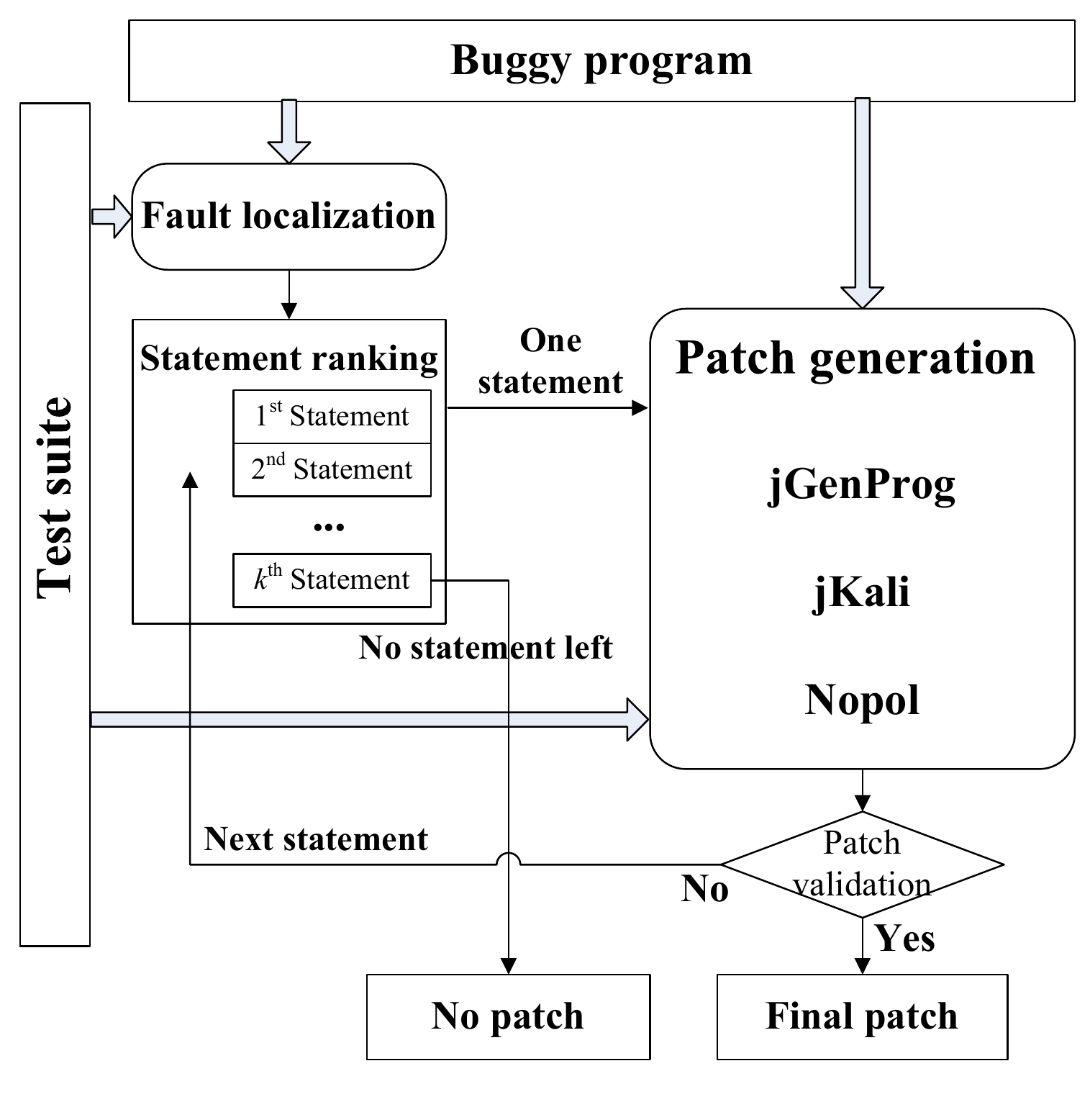}
\caption{Overview of test-suite based repair. Automatic repair takes a buggy program and its test suite as input; the output is the patch that passes the whole test suite if any.}
\label{fig:framework}
\end{figure}

\section{Background}
\label{sect:background}

In this paper, we consider one kind of automatic repair called test-suite based repair.
We now give the corresponding background and present the dataset that is used in our experiment. 

\subsection{Overview of Automatic Repair Techniques}
\label{subsect:automatic-repair}

We briefly give an overview of automatic repair. For a survey on the field, we refer the reader to \cite{Monperrus2015}.
There are two families of repair techniques: offline repair consists of generating source code patches, online repair, aka runtime repair, consists of modifying the system state at runtime to overcome failures. The latter is the direct descendant of classical fault tolerance. 
Automatic repair techniques use a variety of oracles to encode the expected behavior of the program under repair: the notable oracles are test-suites, pre- and post-conditions, formal behavioral models and runtime assertions.
In the experiment presented in this paper, the considered benchmark makes us considering \emph{offline patch generation} using \emph{test suites} as oracles.

\subsection{Test-Suite Based Repair}
\label{subsect:test-suite-based-repair}

Test-suite based repair generates a patch according to failing test cases. Different kinds of techniques can be used, such as genetic programming search in GenProg (\cite{le2012genprog}) and SMT based program synthesis in SemFix (\cite{nguyen2013semfix}). Often, before patch generation, a fault localization method is applied to rank the statements according to their suspiciousness. The intuition is that the patch generation technique is more likely to be successful on suspicious statements. 

Fig. \ref{fig:framework} presents a general overview of test-suite based repair approaches. In a repair approach, the input is a buggy program as well as its test suite; the output is a patch that makes the test suite pass, if any. To generate a patch for the buggy program, the executed statements are ranked to identify the most suspicious statements. \textit{Fault localization} is a family of techniques for ranking potential buggy statements (\cite{jones2002visualization,abreu2007accuracy,xuan2014test}). Based on the statement ranking, \textit{patch generation} tries to modify a suspicious statement. For instance, GenProg (\cite{le2012genprog}) adds, removes, and replaces Abstract Syntax Tree (AST) nodes. Once a patch is found, the whole test suite is executed to validate the patch; if the patch is not validated by the test suite, the repair approach goes on with next statement and repeats the repair process.

\subsection{Defects4J Dataset}
\label{sec:defects4j}

Defects4J by \cite{JustJE2014} is a bug database that consists of 357 real-world bugs from five widely-used open-source Java projects. Bugs in Defects4J are organized in a unified structure that abstracts over programs, test cases, and patches. 

Defects4J provides research with reproducible software bugs and enables controlled studies in software testing research (e.g., \cite{noor2015test,just2014mutants}). To our knowledge, Defects4J is the largest open database of well-organized real-world Java bugs. In our work, we use four out of five projects, i.e., Commons Lang,\footnote{Apache Commons Lang, \url{http://commons.apache.org/lang}.} JFreeChart,\footnote{JFreeChart, \url{http://jfree.org/jfreechart/}.} Commons Math,\footnote{Apache Commons Math, \url{http://commons.apache.org/math}.} and Joda-Time.\footnote{Joda-Time, \url{http://joda.org/joda-time/}.} We do not use the Closure Compiler project\footnote{Google Closure Compiler, \url{http://code.google.com/closure/compiler/}.} because the test cases in Closure Compiler are organized in a non-conventional way, using scripts rather than standard JUnit test cases. This prevents us from running with our platform and is left for future work.
Table \ref{tab:defects4j} presents the main descriptive statistics of bugs in Defects4J. 

\emph{The advantages of using Defects4J for a automatic repair experiment are:
(realism) it contains real bugs (as opposed to seeded bugs as in \cite{nguyen2013semfix,kong2015experience});
(scale)  bugs are in large software (as opposed to bugs in student programs as in \cite{Smith15fse});
(novelty) nobody has ever studied Defects4J for repair.
}

\begin{table}[!t] 
\caption{The Main Descriptive Statistics of Considered Bugs in Defects4J. The Number of Lines of Code and the Number of Test Cases are Extracted from the Most Recent Version of Each Project.}
\label{tab:defects4j}
\centering
\begin{tabular}{|c|c|c|c|c|}

\hline
Project & \#Bugs & Source KLoC & Test KLoC & \#Test cases \\ \hline\hline
Commons Lang & 65 &  22 & 6 & 2,245 \\ 
JFreeChart & 26 & 96 & 50 & 2,205 \\%Includes C8
Commons Math & 106 & 85 & 19 & 3,602 \\%includes M99
Joda-Time & 27 & 28 & 53 & 4,130  \\
\hline  \hline
Total & \nbAllBugs & 231 & 128 & 12,182 \\
\hline

\end{tabular}
\end{table}

\section{Experimental Protocol}
\label{sect:protocol}

We present an experimental protocol to assess the effectiveness of different automatic repair approaches on the real-world bugs of Defects4J.
The protocol supports the analysis of several dimensions of automatic repair: test-suite adequacy, patch correctness, under-specified bugs, performance.
We first list the Research Questions (RQs) of our work; then we describe the experiment design; finally, we present the implementation details. 

\subsection{Research Questions}
\label{subsect:rq}

\subsubsection{\textbf{RQ1}. Test-suite Adequate Repair} 
For how many bugs can each system synthesize a patch that is test-suite adequate, i.e., it fulfills the test suite?

This is the basic evaluation criterion of automatic test-suite based repair research. 
In test-suite based repair, a bug is said to be automatically patchable if the synthesized patch makes the whole test suite pass. 
Test-suite adequacy does not convey a notion a correctness beyond the test suite. 
To answer this question, we run each repair approach on each bug and count the number of bugs that are patched according the test suite. 

\subsubsection{\textbf{RQ2}. Patch Correctness} 
Which generated patches are semantically correct beyond passing the test suite? 

A patch that passes the whole test suite may not be exactly the same as the patch written by developers. 
It may be syntactically different yet correct.
As shown by \cite{qi2015efficient} it may also be incorrect. 
To answer this question, we manually examine all synthesized patches to identify the correctness.

\subsubsection{\textbf{RQ3}. Under-Specified Bugs} 
Which bugs in Defects4j are not sufficiently specified by the test suite?  

In test-suite based repair, the quality of a synthesized patch is highly dependent on the quality of the test suite. 
In this paper, we define an ``under-specified bug'' as a bug for which the test cases that specify the expected behavior have a low coverage and weak assertions.
Among the incorrect patches identified in RQ2, we identify the bugs for which the primary reason for incorrectness is the test suite itself (and not the repair technique).
To find such under-specified bugs, we use two pieces of evidence.
First, we closely look at the results of \jkali.
Since this repair system only removes code or skips code execution, if it finds a patch, it hints that a functionality is not specified at all.
Second, we complement this by a manual analysis of generated patches. 

\subsubsection{\textbf{RQ4}. Performance (Execution Time)} 
How long is the execution time of each repair system? 

It is time-consuming to manually repair a bug. Test-suite based repair automates the process of patch generation. To assess to which extent automatic repair is usable in practice, we evaluate the execution time of each repair approach.

\subsection{Experiment Design}

According to the inclusion criteria presented in Section \ref{sec:inclusion-criterion}, we select three repair systems (Section \ref{sec:selected-techniques}) on the Defects4J dataset (Section \ref{sec:defects4j}).
Since the experiment requires a large amount of computation, we run it on a grid (Section \ref{sec:grid5000}).
We then manually analyze all the synthesized patches (Section \ref{sec:manual}). 

\subsubsection{Inclusion Criteria}
\label{sec:inclusion-criterion}

For selecting repair systems, we have three inclusion criteria.

\textbf{Criterion \#1}
In this experiment, we consider a dataset of bugs in software written in the Java programming language.
We consider repair systems that are able to handle this programming language. This rules out many systems that repair C code, such as GenProg (\cite{le2012genprog}) and Semfix (\cite{nguyen2013semfix}).

\textbf{Criterion \#2}
Defects4J contains bugs written in modern Java versions (Java 5, 7 and 8). We consider repair systems that support those versions. This rules out Arcuri's pioneering prototype Jaff (\cite{DBLP:conf/cec/ArcuriY08}).

\textbf{Criterion \#3}
The third inclusion criterion is that the repair system is available, whether publicly or on demand. This rules out for instance PAR (\cite{Kim2013}).

\subsubsection{Repair Systems Under Study}
\label{sec:selected-techniques}

As of submission time, there are three repair systems that meet all those criteria:
jGenProg (\cite{astorcode,AstorPaper2016}), jKali (\cite{astorcode}) and Nopol (\cite{nopolcode}).

\paragraph{jGenProg}
jGenProg is an implementation of GenProg for Java.
It implements the GenProg algorithm (\cite{le2012genprog}) as follows. It randomly deletes, adds, and replaces AST nodes in the program. The modification point is steered by spectrum based fault localization. Pieces of code that are inserted through addition or replacement always come from the same program, based on the ``redundancy hypothesis'' (\cite{DBLP:conf/icse/MartinezWM14}). 
There is a threat that this Java implementation does not reflect the actual performance of the original GenProg system for C, it is discussed in Section \ref{threat:reimplementation}.
We are the authors of jGenProg, and we make it publicly available at Github \cite{astorcode,AstorPaper2016}.
Our rationale for implementing jGenProg is that GenProg (which works for C code) is arguably the baseline of this research area, being the most known and visible patch generation system. 
jGenProg has 9.5K lines of Java code.

\paragraph{jKali}
jKali is an implementation of Kali for Java.
Kali (\cite{qi2015efficient}) performs patch synthesis by only removing or skipping code. 
Kali is clearly not a ``program repair'' technique. However, it is a perfectly appropriate technique  to identify weak test suites and under-specified bugs. 
We are the authors of jKali and we make it publicly available (\cite{astorcode}).
Our rationale for implementing Kali is that is an appropriate technique to detect test suite weaknesses, yet the original Kali works for C. 

\paragraph{Nopol}
Nopol is a repair system for Java (\cite{demarco2014automatic,nopoljournal}) which targets a specific fault class: conditional bugs. It repairs programs by either modifying an existing if-condition or  adding a precondition (aka. a guard) to any statement or block in the code. The modified or inserted condition is synthesized via input-output based code synthesis with SMT solvers (\cite{jha2010oracle}).
Nopol is publicly available (\cite{nopolcode}) and has 25K lines of Java code. 

\paragraph{Open Science}
We note that we are authors of those three systems. This is both good and bad. 
This is good because we know that they are of similar quality. 
This is bad because it shows that as of today, although a number of automatic repair contributions have been made for Java, the current practice is not to share the tools for sake of open and reproducible science.

\subsubsection{Large Scale Execution}
\label{sec:grid5000}
We assess three repair approaches on \nbAllBugs bugs. One repair attempt may take hours to be completed. 
Hence, we need a large amount of computation power.
Consequently, we deploy our experiment in Grid'5000, a grid for high performance computing (\cite{grid5000}).
In our experiment, we manually set the computing nodes in Grid'5000 to the same hardware architecture. This avoids potential biases of the time cost measurement. 
All the experiments are deployed in the Nancy site of Grid'5000 (located in Nancy, France). 

As we shall see later, this experiment is significantly CPU intensive: it takes in total \experimenttimeindays days of computation on Grid'5000. This means we had to make careful choices before coming to valid run of the experiment.
In particular, we set the timeout to three hours per repair attempt, in order to have an upper bound. 

For the same reason, although jGenProg is a randomized algorithm, for all but 5 bugs, we run it only once also to keep an acceptable experiment time,  this important threat to validity is discussed in Section \ref{threat:random}. This results in a conservative under-estimation of GenProg's effectiveness but remains sound (found patches are actually test-suite adequate patches).  
Finally, although our and other's experience with repair has shown that multiple patches exist for the same bug, we stop the execution of a repair attempt after finding the first patch.

\subsubsection{Manual Analysis}
\label{sec:manual}

We manually examine the generated patches. 
For each patch, one of the authors, called thereafter an ``analyst'', analyzed the patch correctness, readability, and the difficulty of validating the correctness (as defined below).
For sake of cross validation, the analyst then presents the results of the analysis to another co-author, called thereafter a ``reviewer'' in a live session. At the end of the live session, when an agreement has been reached between the analyst and the reviewer, a short explanatory text is written. Those  texts are made publicly available for peer review (\cite{githubresults}). 
During agreement sessions, analysts and reviewers applied a conservative patch evaluation strategy: when there were a doubt or a disagreement about the correctness of a patch, is was labeled as ``unknown''. 

\paragraph{Correctness Analysis}
The \textit{correctness} of a patch can be correct, incorrect, or unknown.
The term ``correct'' denotes that a patch is exactly the same or semantically equivalent to the patch that is written by developers.
We assume that all patches included in the Defects4J dataset are correct.
The equivalence is assessed according to the analyst's understanding of the patch.
Analyzing one patch requires a period between a couple of minutes and several hours of work, depending on the complexity of the synthesized patch. On the one hand, a patch that is identical to the one written by developers is obviously true; on the other hand, several patches require a domain expertise that none of the authors has.

Note that in the whole history of automatic program repair, this is only the second time that such a large scale analysis of correctness is being performed (after \cite{qi2015efficient}). While other published literature has studied aspects of the readability and usefulness of generated patches (e.g. \cite{Fry2012,TaoKKX14}), it was on a smaller scale with no clear focus on correctness.

\paragraph{Readability Analysis}
The \textit{readability} of the patch can be ``easy'', ``medium'', or ``hard''; and it results from the analyst opinion on the length and complexity of the patch.
This subjective evaluation concerns:
1) The number of variables involved in the patch: a patch that refers to one or two variables is much easier to understand than a patch that refers to 5 variables.
2) The number of logical clauses: in a condition, a patch with a single conjunction (hence with two logical clauses) is considered easier a patch that with many sub-conjunctions.
3) The number of method calls: when a patch contains a method call, one must understand the meaning and the goal of this call to understand the patch. Hence, the more method calls, the harder the patch to understand.

\paragraph{Difficulty Analysis}
The \textit{difficulty} analysis relates to the difficulty of understanding the problem domain and assessing the correctness.
It indicates the effort that the analyst had to carry out for understanding the essence of the bug, the human patch and the generated patch. 
The analysts agree on a subjective difficulty on the scale ``easy'', ``medium'', ``hard'', or ``expert'' as follows.
``Easy'' means that it is enough to examine the source code of the patch for determining its correctness.
``Medium'' means that one has to fully understand the test case.
``Hard'' means that the analyst has to dynamically debug the buggy and/or the patched application. 
By ``expert'', we mean a patch for which it was impossible for us to validate  due to the required expertise in domain knowledge. This happens when the analyst could not understand the bug, or the developer solution or the synthesized patch.

\section{Empirical Results}
\label{sect:result}

We present and discuss our answers to the research questions that guide this work.
The total execution of the experiment costs \experimenttimeindays days.

\begin{table}[!t]
\caption{Results for \nbAllBugs Bugs in Defects4J with Three Repair Approaches. In Total, the Three Repair Approaches can Generate Test-suite Adequate Patches for \nbFixedBugs Bugs (21\%).
}
\label{tab:bug-fix}
\centering

\resizebox{0.45\textwidth}{!}
{
\setlength\tabcolsep{0.7 ex}
\begin{tabular}{|c|c|c|c|c|}
\hline 
Project & Bug Id & jGenProg & jKali & Nopol   \\
\hline\hline
\multirow{9}{*}{ \rotatebox{90}{JFreeChart}}  
&	C1                	&	 Patch     	&	 Patch     	&	 --        	\\
&	C3                	&	 Patch     	&	 --        	&	 Patch     	\\
&	C5                	&	 Patch     	&	 Patch     	&	 Patch     	\\
&	C7                	&	 Patch     	&	 --        	&	 --        	\\
&	C13               	&	 Patch     	&	 Patch     	&	 Patch     	\\
&	C15               	&	 Patch     	&	 Patch     	&	 --        	\\
&	C21               	&	 --     	  &	 --     	  &	 Patch     	\\
&	C25               	&	 Patch     	&	 Patch     	&	 Patch     	\\
&	C26               	&	 --        	&	 Patch     	&	 Patch     	\\
\hline
\multirow{7}{*}{ \rotatebox{90}{Commons Lang}} 
&	L39               	&	 --        	&	 --        	&	 Patch	\\
&	L44               	&	 --        	&	 --        	&	 Patch     	\\
&	L46               	&	 --        	&	 --        	&	 Patch	\\
&	L51               	&	 --        	&	 --        	&	 Patch     	\\
&	L53               	&	 --        	&	 --        	&	 Patch	\\
&	L55               	&	 --        	&	 --        	&	 Patch	\\
&	L58               	&	 --        	&	 --        	&	 Patch     	\\
\hline
\multirow{2}{*}{\rotatebox{90}{\small Time}}
&	T4                	&	 Patch     	&	 Patch     	&	 --        	\\
&	T11               	&	 Patch     	&	 Patch     	&	 Patch     	\\
\hline
&	M2                	&	 Patch     	&	 Patch     	&	 --        	\\
&	M5                	&	 Patch     	&	 --        	&	 --        	\\
&	M8                	&	 Patch     	&	 Patch     	&	 --        	\\
&	M28               	&	 Patch     	&	 Patch     	&	 --        	\\
&	M32               	&	 --        	&	 Patch     	&	 Patch	\\

\hline 
\end{tabular}
}
\resizebox{0.51\textwidth}{!}{
\setlength\tabcolsep{0.7 ex}
\begin{tabular}{|c|c|c|c|c|}
\hline 
Project & Bug Id & jGenProg & jKali & Nopol   \\
\hline\hline
\multirow{20}{*}{ \rotatebox{90}{Commons Math}} 
   	&	M33               	&	 --        	&	 --        	&	 Patch  \\
&	M40               	&	 Patch     	&	 Patch     	&	 Patch     	\\
&	M42               	&	 --        	&	 --        	&	 Patch	\\
&	M49               	&	 Patch     	&	 Patch     	&	 Patch     	\\
&	M50               	&	 Patch     	&	 Patch     	&	 Patch     	\\
&	M53               	&	 Patch     	&	 --        	&	 --        	\\
&	M57               	&	 --        	&	 --        	&	 Patch     	\\
&	M58               	&	 --        	&	 --        	&	 Patch     	\\
&	M69               	&	 --        	&	 --        	&	 Patch	\\
&	M70               	&	 Patch     	&	 --        	&	 --        	\\
&	M71               	&	 Patch     	&	 --        	&	 Patch     	\\
&	M73               	&	 Patch     	&	 --        	&	 Patch     	\\
&	M78               	&	 Patch     	&	 Patch     	&	 Patch	\\
&	M80               	&	 Patch     	&	 Patch     	&	 Patch	\\
&	M81               	&	 Patch     	&	 Patch     	&	 Patch     	\\
&	M82               	&	 Patch     	&	 Patch     	&	 Patch	\\
&	M84               	&	 Patch     	&	 Patch     	&	 --        	\\
&	M85               	&	 Patch     	&	 Patch     	&	 Patch     	\\
&	M87               	&	 --        	&	 --        	&	 Patch     	\\
&	M88               	&	 --        	&	 --        	&	 Patch	\\
&	M95               	&	 Patch     	&	 Patch     	&	 --        	\\
&	M97               	&	 --        	&	 --        	&	 Patch     	\\
&	M104              	&	 --        	&	 --        	&	 Patch     	\\
&	M105              	&	 --        	&	 --        	&	 Patch	\\
\hline
\hline
\hline
Total & 47 (21\%) & 27 (12\%) & 22 (9.8\%) & 35 (15.6\%) \\
\hline 
\end{tabular}

}

\end{table}

\subsection{Test-suite Adequate Repair} 
\label{subsect:answer-rq1}

\textbf{RQ1}. For how many bugs can each system synthesize a patch that fulfills the test suite?

The three automatic repair approaches in this experiment are able to together find test-suite adequate patches for \nbFixedBugs bugs of the Defects4J dataset. 
\jgenprog finds a patch for \nbFixedBugsByGenprog bugs; \jkali identifies a patch for \nbFixedBugsByKali bugs; and Nopol synthesizes a condition that makes the test suite passing for \nbFixedBugsByNopol bugs. 
Table \ref{tab:bug-fix} shows the bug identifiers, for which at least one patch is found.
Each line corresponds to one bug in Defects4J and each column denotes the effectiveness of one repair approach. 
For instance, \jgenprog and \jkali are able to synthesize a test-suite adequate patch for Bug M2 from Commons Math.

As shown in Table \ref{tab:bug-fix},
some bugs such as T11 can be patched by all systems, others by only a single one.
For instance, only Nopol synthesize a test-suite adequate patch for bug  L39 and only \jgenprog for bug M5.

Moreover, Table \ref{tab:bug-fix} shows that in project Commons Lang only Nopol generates test-suite adequate patches while \jgenprog and \jkali fail to synthesize a single patch. A possible reason is that the program of Commons Lang is more complex than that of Commons Math; both \jgenprog and \jkali cannot handle such a complex search space.

\begin{figure}[!t]
\centering
\includegraphics[width=0.5\textwidth]{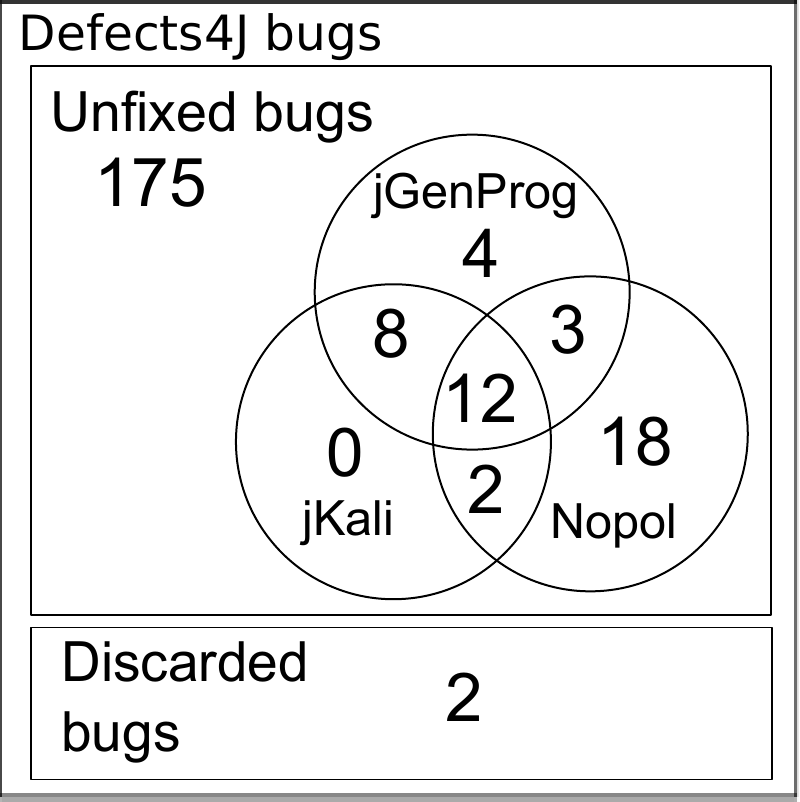}
\caption{Venn diagram of bugs for which test-suite adequate patches are found.}
\label{fig:intersection}
\end{figure}

Fig. \ref{fig:intersection} shows the intersections among the three repair approaches as a Venn diagram, where each number is a number of patches for which one system is able to generate a test-suite adequate patch.
Nopol can synthesize a patch for 18 bugs that neither \jgenprog nor \jkali could handle. All the bugs handled by \jkali can also be handled by \jgenprog and Nopol. For \nbFixedBugsByAll bugs, all three repair systems can generate a patch to pass the test suite.  
The 2 discarded bugs will be discussed in Section \ref{sec:falky}.

To our knowledge, those results are the very first on automatic repair  with the Defects4J benchmark. Recall that they are done with an open-science ethics, all the implementations, experimental code, and results are available on Github (\cite{githubresults}). 
Future research in automatic repair may try to synthesize more test-suite adequate patches than our work. Our experimental framework can be used to facilitate future comparisons by other researchers.   

Among the incorrect patches identified in RQ2, we identify the bugs for which the primary reason for incorrectness is the test suite itself (and not the repair technique).
\\

\noindent\fbox{\parbox{\textwidth}{
\textbf{Answer to RQ1}. In Defects4J, at least one of the automatic repair systems under consideration can generate a test-suite adequate patch 
for \nbFixedBugs out of \nbAllBugs bugs. Nopol can is the most effective (\nbFixedBugsByNopol bugs). If \jkali finds a  test-suite adequate patch, so can \jgenprog or Nopol. This experiment sets a record to be beaten by future yet-to-be-invented repair systems.
}}

\subsection{Patch Correctness}
\label{subsect:answer-rq2}

\textbf{RQ2}. Which patches are semantically correct beyond passing the test suite?  

\begin{landscape}
\begin{table}[!t]

\caption{Manual Assessment of \nbDifferentPatches Patches that are Generated by Three Repair Approaches.}
\label{tab:bug-correct}

\resizebox{0.43\linewidth}{!}
{
\begin{tabular}{|c|c|c|c|c|c|c|}
\hline 
Project & Bug id & Patch id & Approach & Correctness    & Readability & Difficulty \\
\hline 
\hline 
\multirow{19}{*}{ \rotatebox{90}{JFreeChart}} 
& C1   & 1  & jGenProg & Incorrect         & Easy       & Easy   \\%a15
& C1   & 2  & jKali    & Incorrect         & Easy       & Easy   \\%a15
& C3   & 3  & jGenProg & Unknown           & Medium     & Medium \\%a15 Can be correct?
& C3   & 4  & Nopol    & Incorrect         & Easy       & Medium \\%a15
& C5   & 5  & jGenProg & Incorrect         & Easy       & Medium \\%a15
& C5   & 6  & jKali    & Incorrect         & Easy       & Medium \\%a15
& C5   & 7  & Nopol    & \textbf{Correct}  & Easy       & Medium \\%a15
& C7   & 8  & jGenProg & Incorrect         & Easy       & Easy   \\%a15
& C13  & 9  & jGenProg & Incorrect         & Easy       & Easy   \\%a15
& C13  & 10 & jKali    & Incorrect         & Easy       & Easy   \\%a15
& C13  & 11 & Nopol    & Incorrect         & Easy       & Easy   \\
& C15  & 12 & jGenProg & Incorrect         & Easy       & Medium \\%a15
& C15  & 13 & jKali    & Incorrect         & Medium     & Medium \\%a15
& C21  & 14 & Nopol    & Incorrect         & Hard       & Expert \\
& C25  & 15 & jGenProg & Incorrect         & Medium     & Medium \\%a15
& C25  & 16 & jKali    & Incorrect         & Medium     & Medium \\%a15
& C25  & 17 & Nopol    & Incorrect         & Easy       & Easy   \\
& C26  & 18 & jKali    & Incorrect         & Easy       & Medium \\%a15
& C26  & 19 & Nopol    & Incorrect         & Easy       & Medium \\%a15
\hline
\multirow{7}{*}{\rotatebox{90}{Commons Lang}} 
& L39  & 20 & Nopol    & Incorrect         & Easy       & Medium \\%a15
& L44  & 21 & Nopol    & \textbf{Correct}  & Easy       & Medium \\
& L46  & 22 & Nopol    & Incorrect         & Easy       & Medium \\
& L51  & 23 & Nopol    & Incorrect         & Easy       & Easy   \\
& L53  & 24 & Nopol    & Incorrect         & Hard       & Expert \\%a15
& L55  & 25 & Nopol    & \textbf{Correct}  & Easy       & Medium \\
& L58  & 26 & Nopol    & \textbf{Correct}  & Easy       & Medium \\
\hline
\multirow{5}{*}{\rotatebox{90}{Time}} & 
  T4   & 80 & jGenProg & Incorrect         & Easy       & Medium \\%August15
& T4   & 81 & jKali    & Incorrect         & Easy       & Medium \\%August1
& T11  & 82 & jGenProg & Incorrect         & Easy       & Easy   \\%a15
& T11  & 83 & jKali    & Incorrect         & Easy       & Easy   \\%a15
& T11  & 84 & Nopol    & Incorrect         & Medium     & Medium \\%a15
\hline
\multirow{11}{*}{ \rotatebox{90}{Commons Math}} & 
 M2   & 27 & jGenProg & Incorrect         & Easy       & Hard   \\
& M2   & 28 & jKali    & Incorrect         & Easy       & Hard   \\
& M5   & 29 & jGenProg & \textbf{Correct}  & Easy       & Easy   \\%a15
& M8   & 30 & jGenProg & Incorrect         & Easy       & Easy   \\
& M8   & 31 & jKali    & Incorrect         & Easy       & Easy   \\
& M28  & 32 & jGenProg & Incorrect         & Medium     & Hard   \\%a15 to discuss
& M28  & 33 & jKali    & Incorrect         & Easy       & Hard   \\%a15 to discuss
& M32  & 34 & jKali    & Incorrect         & Easy       & Easy   \\%a15 to discuss
& M32  & 35 & Nopol    & Unknown           & Hard       & Expert \\
& M33  & 36 & Nopol    & Incorrect         & Medium     & Medium \\
& M40  & 37 & jGenProg & Incorrect         & Hard       & Hard   \\%a15
& M40  & 38 & jKali    & Incorrect         & Easy       & Medium \\%a15
\hline
\end{tabular}
}
\resizebox{0.47\linewidth}{!}
{
\begin{tabular}{|c|c|c|c|c|c|c|}
\hline 
Project & Bug id & Patch id & Approach & Correctness    & Readability & Difficulty \\
\hline 
\hline
\multirow{41}{*}{ \rotatebox{90}{Commons Math}} 
& M40  & 39 & Nopol    & Unknown           & Hard       & Expert \\%a15
& M42  & 40 & Nopol    & Unknown           & Medium     & Expert \\
& M49  & 41 & jGenProg & Incorrect         & Easy       & Medium \\
& M49  & 42 & jKali    & Incorrect         & Easy       & Medium \\
& M49  & 43 & Nopol    & Incorrect         & Easy       & Medium \\
& M50  & 44 & jGenProg & \textbf{Correct}  & Easy       & Easy   \\
& M50  & 45 & jKali    & \textbf{Correct}  & Easy       & Easy   \\
& M50  & 46 & Nopol    & \textbf{Correct}  & Easy       & Medium \\
& M53  & 47 & jGenProg & \textbf{Correct}  & Easy       & Easy   \\%a15
& M57  & 48 & Nopol    & Incorrect         & Medium     & Medium \\
& M58  & 49 & Nopol    & Incorrect         & Medium     & Hard   \\
& M69  & 50 & Nopol    & Unknown           & Medium     & Expert \\
& M70  & 51 & jGenProg & \textbf{Correct}  & Easy       & Easy   \\
& M71  & 52 & jGenProg & Unknown           & Medium     & Hard   \\%a15
& M71  & 53 & Nopol    & Incorrect         & Medium     & Hard   \\
& M73  & 54 & jGenProg & \textbf{Correct}  & Easy       & Easy   \\
& M73  & 55 & Nopol    & Incorrect         & Easy       & Easy   \\
& M78  & 56 & jGenProg & Unknown           & Easy       & Hard   \\
& M78  & 57 & jKali    & Unknown           & Easy       & Hard   \\
& M78  & 58 & Nopol    & Incorrect         & Medium     & Hard   \\
& M80  & 59 & jGenProg & Incorrect         & Hard       & Medium \\
& M80  & 60 & jKali    & Unknown           & Easy       & Medium \\
& M80  & 61 & Nopol    & Unknown           & Easy       & Medium \\
& M81  & 62 & jGenProg & Incorrect         & Easy       & Medium \\
& M81  & 63 & jKali    & Incorrect         & Easy       & Medium \\
& M81  & 64 & Nopol    & Incorrect         & Easy       & Medium \\
& M82  & 65 & jGenProg & Incorrect         & Easy       & Medium \\
& M82  & 66 & jKali    & Incorrect         & Easy       & Medium \\
& M82  & 67 & Nopol    & Incorrect         & Easy       & Medium \\
& M84  & 68 & jGenProg & Incorrect         & Easy       & Easy   \\
& M84  & 69 & jKali    & Incorrect         & Easy       & Easy   \\     
& M85  & 70 & jGenProg & Unknown           & Easy       & Easy   \\
& M85  & 71 & jKali    & Unknown           & Easy       & Easy   \\
& M85  & 72 & Nopol    & Incorrect         & Easy       & Easy   \\
& M87  & 73 & Nopol    & Incorrect         & Medium     & Expert \\%a15
& M88  & 74 & Nopol    & Incorrect         & Easy       & Medium \\
& M95  & 75 & jGenProg & Incorrect         & Easy       & Hard   \\
& M95  & 76 & jKali    & Incorrect         & Easy       & Hard   \\
& M97  & 77 & Nopol    & Incorrect         & Easy       & Medium \\%a15
& M104 & 78 & Nopol    & Incorrect         & Hard       & Expert \\%a15
& M105 & 79 & Nopol    & Incorrect         & Medium     & Medium \\%a15
\hline  
\hline
\multicolumn{4}{|c|}{ \nbDifferentPatches Patches for \nbFixedBugs bugs}        & \nbAnalyzedPatchCorrect Correct        & \nbAnalyzedReaEasy Easy    & \nbAnalyzedDiffHardExpert Hard/Expert  \\
\hline
\multicolumn{7}{|c|}{ 5 patches correct from jGenProg, 1 from jKali and 5 from Nopol}\\
\hline 
\end{tabular}
}

\end{table}

\end{landscape}

\begin{figure}[!t]
\centering
\includegraphics[width=0.8\textwidth]{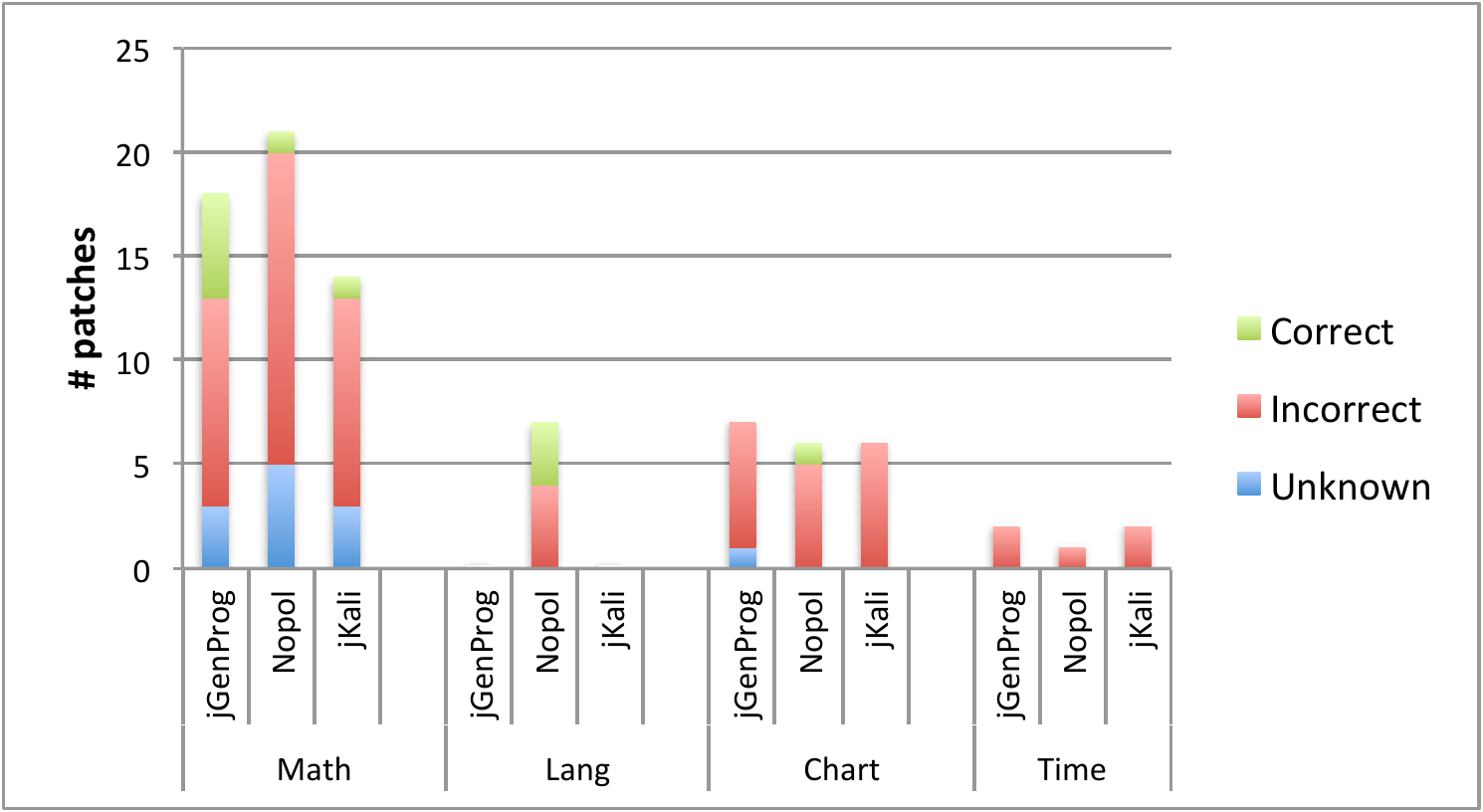}
\caption{Bar chart that summarizes the results from the correctness analysis.}
\label{fig:patch_correctness}
\end{figure}

We manually evaluate the correctness of generated patches by the three repair approaches under study. A generated patch is \textit{correct} if this patch is the same as the manually-written patch by developers or the patch is semantically equivalent to the manual patch. A generated patch is \textit{incorrect} if it actually does not repair the bug (beyond test-suite adequacy, beyond making the failing test case pass -- a kind of incomplete bug oracle) or if it breaks an expected behavior (beyond keeping the rest of the test suite passing).

Recall the history of automatic repair research.
It has been hypothesized that a major pitfall of test-suite based repair is that a test suite cannot completely express the program specifications, so it is hazardous to drive the synthesis of a correct patch with a test suite (\cite{qi2015efficient,Fry2012}). 
Previous works have studied the maintainability of automatic  generated patches (\cite{Fry2012}) or their aids for debugging task (\cite{TaoKKX14}).
However, only recent work by \cite{qi2015efficient} has invested resources to manually analyze the correctness previously-generated patches by test-suite based repair. They found that the vast majority of patches by GenProg in the GenProg benchmark of C bugs are incorrect.

To answer the question of patch correctness, we have manually analyzed all the patches generated by Nopol, \jgenprog and \jkali in our experiment, \nbDifferentPatches patches in total. 
This represents more than ten full days of work. To our knowledge, only \cite{qi2015efficient} have performed a similar manual assessment of patches synthesized with automatic repair. 

Table \ref{tab:bug-correct} shows the results of this manual analysis. The ``bug id'' column refers to the Defects4J identifier, while ``Patch id'' is an univocal identifier of each patch, for easily identifying the patch on our empirical result page (\cite{githubresults}).
The three main columns give the correctness, readability and difficulty as explained in Section \ref{sec:manual}. 

The results in Table \ref{tab:bug-correct} may be fallible due to the subjective nature of the assessment. For mitigating this risk, all patches as well as detailed case studies are publicly available on Github (\cite{githubresults}).  

In total, we have analyzed \nbDifferentPatches patches. Among these patches, \nbFixedBugsByGenprog, \nbFixedBugsByKali, and \nbFixedBugsByNopol patches are synthesized by \jgenprog, \jkali, and Nopol, respectively.

As shown in Table \ref{tab:bug-correct}, \nbAnalyzedPatchCorrect out of \nbDifferentPatches analyzed patches are correct and \nbAnalyzedIncorrect are incorrect. Meanwhile, for the other \nbAnalyzedUnknown patches, it is not possible to clearly validate the correctness, due to the lack of domain expertise (labeled as \textit{unknown}). 
Figure \ref{fig:patch_correctness} graphically summarizes this analysis. This figure clearly convey the fact that most test-suite adequate patches are actually incorrect because of overfitting.
Section \ref{sect:case-study} will present three case studies of generated patches via manual analysis.

Among the \nbAnalyzedPatchCorrect correct patches, \jgenprog, \jkali, and Nopol contribute to 5, 1, and 5 patches, respectively. All the correct patches by \jgenprog and \jkali come from Commons Math; 3 correct patches by Nopol come from Commons lang, one comes from JFreeChart and the other from Commons Math. After the controversy about the effectiveness of GenProg (\cite{qi2015efficient}), it is notable to see that there are bugs for which only \jgenprog works for real.

Among \nbDifferentPatches analyzed patches, \nbAnalyzedReaEasy patches are identified as easy to read and understand.
For the difficulty of patch validation, \nbAnalyzedDiffHardExpert patches are labeled as hard or expert. This result shows that it is hard and time consuming to conduct a manual assessment of patches. 

For the incorrect patches, the main reasons are as follows.
First, all three approaches are able to remove some code (pure removal for \jkali, replacement for \jgenprog, precondition addition for Nopol). The corresponding patches simply exploit some under-specification and remove the faulty but otherwise not tested behavior. 
When the expected behavior seems to be well-specified (according to our understanding of the domain), the incorrect patches tend to overfit the test data. For instance, if a failing test case handles a $2 \times 2$ matrix, 
the patch may use such test data to incorrectly force the patch to be suitable only for matrices of size of $2 \times 2$. 

This goes along the line of \cite{qi2015efficient}'s results  and \cite{Smith15fse}' findings. Compared to \cite{Smith15fse}, we study real bugs in large programs and not small ones in student programs. Compared to \cite{qi2015efficient}, we study a different benchmark in a different programming language.
To this extent, the key novelty of our experimental results is that:
1) overfitting also happens for Java software tested with a state-of-the-art testing infrastructure (as opposed to old-school C programs in \cite{qi2015efficient});
1) overfitting also happens with Nopol, a synthesis based technique (and not only for Genprog and similar techniques).
Our results clearly identify overfitting as the next big challenge of test-suite based automatic repair.
\\

\noindent\fbox{\parbox{\textwidth}{
\textbf{Answer to RQ2}. Based on the manual assessment of patch correctness, we find out that only \nbAnalyzedPatchCorrect out of \nbDifferentPatches generated patches are semantically correct (beyond passing the provided test suite).  
The reason is that the current test suite based repair algorithms tend to overfit the test case data or to exploit the holes left by insufficiently specified functionality.}}

\subsection{Under-specified bugs}
\label{subsect:answer-rq3}
\textbf{RQ3}. Which bugs in Defects4j are not sufficiently specified by the test suite?

As shown in Section \ref{subsect:answer-rq1}, the repair system \jkali can find test-suite adequate patches for \nbFixedBugsByKali bugs. Among these generated patches, from our manual evaluation, we find out that 18 patches are incorrect (other 3 patches are unknown). In each of those generated patches by \jkali, one statement is removed or skipped to eliminate the failing program behavior, instead of making it correct. 
This kind of patches shows that the corresponding test suite is too weak with respect to the buggy functionality. 
The assertions that specify the expected behavior of the removed statement and the surrounding code are inexistent or too weak. 

One exception among \nbFixedBugsByKali patches by \jkali is the patch of Bug M50. As shown in Section  \ref{subsect:answer-rq2}, the patch of Bug M50 is correct. That is, the statement removal is the correct patch. 
Another special case is Bug C5 which is patched by \jkali (incorrect) and by Nopol (correct). The latter approach produces a patch similar to that done by the developer. 
A patch (written by developers or automatically generated) that is test-suite adequate for an under-specified bug could introduce new bugs (studied previously by \cite{Zhongxian2010buggyfix}) or it could not be completely correct due to a weak test suite used as the bug oracle (\cite{qi2015efficient}).
Table \ref{tab:bug-remove} summarizes this finding and list the under-specified bugs.

This result is important for future research on automatic repair with Defects4J.
First, any repair system that claims to correctly repair one of those bugs should be validated with a detailed manual analysis of patch correctness, to check whether the patch is not a variation on the trivial removal solution.
Second, those bugs can be considered as the most challenging ones of Defects4J. To repair them, a repair system must somehow reason about the expected functionality beyond what is encoded in the test suite.
This is what was actually been done by the human developer. 
A repair system that is able to produce a correct patch for those bugs would be a great advance for the field.

\noindent\fbox{\parbox{\textwidth}{
\textbf{Answer to RQ3}. Among the incorrect patches identified in RQ2, there are 21 bugs for which the primary reason for incorrectness is the test suite itself (and not the repair technique), they are under-specified bugs. For them, the test suite does not accurately specify the expected behavior and can be trivially repaired by removing code.
To us, they are the most challenging bugs: to automatically repair them, one needs to reason on the expected functionality below what is encoded in the test suite for instance by taking into account a source of information other than test suite execution. Repairing those bugs can be seen as the next grand challenge for automatic repair.}}

\begin{table}[!t] 
\caption{The Most Challenging Bugs of Defects4J Because of Under-specification.}
\label{tab:bug-remove}
\centering
\begin{tabular}{|c|c|}

\hline
Project      & Bug ID    \\ \hline\hline
Commons Math & M2, M8,M28,M32,M40, M49, M78, \\ 
	    	     & M80, M81, M82,M84, M85,M95 \\  \hline
JFreeChart & C1,C5, C13, C15, C25,C26\\\hline
Time &T4,T11\\\hline
\end{tabular}
\end{table}

\subsection{Performance}
\label{subsect:answer-rq4}
\textbf{RQ4}. How long is the execution time for each repair approach on one bug? 

For real applicability in industry, automatic repair approaches must execute fast enough. 
By ``fast enough'', we mean an acceptable time period, which depends on the usage scenario of automatic repair and on the hardware. For instance, if automatic repair is meant to be done in the IDE, repair time should last at most some minutes on a standard desktop machine. 
On the contrary, if automatic repair is meant to be done  on a continuous integration server, it is acceptable to last hours on a more powerful server hardware configuration. 

The experiments in this paper are run on a grid where most of nodes have comparable characteristics. 
Typically, we use machines with Intel Xeon X3440 Quad-core processor and 15GB RAM.
Table \ref{tab:bug-time} shows the time cost of patch generation in hours for bugs without timeout. As shown in Table \ref{tab:bug-time}, the median time for one bug by \jgenprog is around one hour.
The fastest repair attempt yields a patch in 31 seconds (for Nopol). 
The median time to synthesize a patch is 6.7 minutes. 
This means that the execution time of automatic repair approaches is comparable to the time of manual repair by developers. It may be even faster, but we don't know the actual repair time by real developers for the bug of the dataset.

When a test-adequate patch can be synthesized, it is found within minutes. This means that most of the time of the \experimenttimeindays days of computation for the experiment is spent on unfixed bugs, which reach the timeout.
For \jgenprog, it is always the case, because the search space is extremely large.
For \jkali, we often completely explore the search space, and we only reach the timeout in 20 cases.
For Nopol, the timeout is reached in 26 cases, either due to the search space of covered statements or the SMT synthesis that becomes slow in certain cases.
One question is whether a larger timeout would improve the effectiveness. 
According to this experiment, the answer is no. The repairability is quite binary: either a patch is found fast, or the patch cannot be found at all. This preliminary observation calls for future research.

\begin{table}[!t] 
\caption{Time Cost of Patch Generation}
\label{tab:bug-time}
\centering
\begin{tabular}{|c|c|c|c|}
\hline
Time cost& \jgenprog  & \jkali       & Nopol        \\\hline\hline
Min    & 40 sec       & 36 sec       & 31sec        \\
Median & 1h 01m       & 18m 45sec    & 22m 30sec      \\
Max    & 1h 16m       & 1h 27m       & 1h 54m       \\\hline
Average& 55m 50sec    & 23m 33sec    & 30m 53sec    \\
Total  & 8 days 12h   & 3 days 6h    & 7 days 3h    \\\hline
\end{tabular}
\end{table}

\noindent\fbox{\parbox{\textwidth}{
\textbf{Answer to RQ4}. Both the median value and the average value of repair execution time are about one hour on  machines that represent server machines used nowadays for continuous integration. Performance is not a problem with respect to practical usage of automatic repair. }}

\subsection{Other Findings in Defects4J}
\label{subsect:other-findings}

Our manual analysis of results enables us to uncover two problems in Defects4J.
First, we found that bug \#8 from project JFreeChart (C8) is flaky, which depends on the machine configuration.
Second, bug \#99 from Commons Math (M99) is identical to bug M97. Both issues were reported to the authors of Defects4J and will be solved in future releases of Defects4J.

\section{Case Studies}
\label{sect:case-study}

In this section, we present three case studies of generated patches by \jgenprog, \jkali, and Nopol, respectively. These case studies are pieces of evidence that:
State-of-the-art automatic repair is able to find correct patches (Sections \ref{subsect:case-m70} and \ref{subsect:case-l55}), but also fails with incorrect patches (Section \ref{subsect:case-m8}).
It is possible to automatically generate the same patch as the one  written by the developer (Section \ref{subsect:case-m70}).

\subsection{Case Study of M70, Bug that is Only Repaired by \jgenprog}
\label{subsect:case-m70}

In this section, we study Bug M70, which is repaired by \jgenprog, but cannot be repaired by \jkali and Nopol.  

Bug M70 in Commons Math is about univariate real function analysis. 
Fig. \ref{fig:case-m70} presents the buggy method of Bug M70. This buggy method contains only one statement, a method call to an overloaded method. In order to perform the correct calculation, the call has to be done with an additional parameter \mycode{UnivariateRealFunction f} (at Line 1) to the method call. Both the manually-written patch and the patch by \jgenprog add the parameter \mycode{f} to the method call (at Line 5). 
This patch generated by \jgenprog is considered correct since the it is the same as that by developers. 

\begin{figure}[!t]
\centering
\noindent\begin{minipage}{0.4\textwidth}
\begin{lstlisting}[numbers=left]
double solve(UnivariateRealFunction f, 
      double min, double max, double initial)
    throws MaxIterationsExceededException, 
      FunctionEvaluationException {
// PATCH: return solve(f, min, max);
  return solve(min, max); 
}
\end{lstlisting}

\end{minipage}
\caption{Code snippet of Bug M70. The manually-written patch and the patch by \jgenprog are the same, which is shown in the \mycode{PATCH} comment at Line 5, which adds a parameter to the method call.} 
\label{fig:case-m70}
\end{figure}

To repair Bug M70, \jgenprog generates a patch by replacing the method call by another one, which is picked elsewhere in the same class. This bug cannot be repaired by either \jkali or Nopol. \jkali removes and skips statements; Nopol only handles bugs that are related to \ourif conditions. Indeed, the fact that certain bugs are only repaired by one tool confirms that the fault classes addressed by each approach are not identical. 
\textbf{To sum up, Bug M7 shows that the GenProg algorithm, as implemented in \jgenprog, is capable of uniquely repairing real Java bugs (only \jgenprog succeeds).}

\subsection{Case Study of M8, Bug that is Incorrectly Repaired by \jkali and \jgenprog}
\label{subsect:case-m8} 

In this section, we present a case study of Bug M8, for which \jkali as well as \jgenprog find a test-suite adequate patch, but not Nopol.  

\begin{figure}[!t]
\centering
\noindent\begin{minipage}{0.4\textwidth}
\begin{lstlisting}[numbers=left]
T[] sample(int sampleSize)  {
  if (sampleSize <= 0) {
    throw new NotStrictlyPositiveException([...]);
  }
// MANUAL PATCH: 
// Object[] out = new Object[sampleSize];
  T[] out = (T[]) Array.newInstance(
    singletons.get(0).getClass(), sampleSize);
  for (int i = 0; i < sampleSize; i++) {
// PATCH: removing the following line
    out[i] = sample();  
  }
  return out;
}
\end{lstlisting}

\end{minipage}
\caption{Code snippet of Bug M8. The manually-written patch is shown in the \mycode{MANUAL PATCH} comment at Lines 5 and 6 (changing a variable type). The patch by \jkali in the \mycode{PATCH} comment removes the loop body at Line 11. }
\label{fig:case-m8}
\end{figure}

Bug M8\footnote{Bug ID in the bug tracking system of Commons Math is Math-942, \url{http://issues.apache.org/jira/browse/MATH-942}.} in Commons Math, is about the failure to create an array of a random sample from a discrete distribution.
Fig. \ref{fig:case-m8} shows an excerpt of the buggy code and the corresponding manual and synthesized patches (from \mycode{class DiscreteDistribution<T>}).
The method \mycode{sample} receives the expected number \mycode{sampleSize} of random values and returns an array of the type \mycode{T[]}. 

The bug is due to an exception thrown at Line 11 during the assignment to \mycode{out[i]}.
The method \mycode{Array.newInstance(class, int)} requires a class of a data type as the first parameter. The bug occurs when 
a) the first parameter is of type \mycode{T1}, which is a sub-class of \mycode{T}  
and 
b) one of the samples is an object which is of type \mycode{T2}, which is a sub-class of \mycode{T}, but not of type \mycode{T1}. 
Due to the incompatibility of types T1 and T2, an \mycode{ArrayStoreException} is thrown when this object is assigned to the array.  

In the manual patch, the developers change the array type in its declaration (from \mycode{T[]} to \mycode{Object[]}) and the way the array is instantiated.
The patch generated by \jkali simply removes the statement, which assigns \mycode{sample()} to the array.
As consequence, method \mycode{sample} never throws an exception but returns an empty array (only containing null values).
This patch passes the failing test case and the full test suite as well.
The reason of this is that the test case has only one assertion: it asserts that the array size is equal to 1. There is no assertion on the content of the returned array.
However, despite passing the test suite, the patch is clearly incorrect. 
This is an example of a bug that is not well specified by the test suite. 
For this bug, \jgenprog can also generate a patch by replacing the assignment by a side-effect free statement, which is semantically equivalent to removing the code. 
\textbf{To sum up, Bug M8 is an archetypal example of  under-specified bugs as detected by the \jkali system.}

\subsection{Case Study of L55, Bug that is Repaired by Nopol, Equivalent to the Manual Patch}
\label{subsect:case-l55} 

Nopol (\cite{demarco2014automatic}) focuses on condition-related bugs. Nopol collects runtime data to synthesize a condition patch. In this section, we present a case study of Bug L55, which is only repaired by Nopol, but cannot be repaired by \jgenprog or \jkali.

\begin{figure}[!t]
\centering
\noindent\begin{minipage}{0.4\textwidth}
\begin{lstlisting}[numbers=left]
void stop() {
  if (this.runningState != STATE_RUNNING 
      && this.runningState != STATE_SUSPENDED) {
    throw new IllegalStateException(...);
  }
// MANUAL PATCH: 
// if (this.runningState == STATE_RUNNING)
// NOPOL PATCH: 
// if (stopTime < StopWatch.STATE_RUNNING)
  stopTime = System.currentTimeMillis();
  this.runningState = STATE_STOPPED;
}
\end{lstlisting}

\end{minipage}
\caption{Code snippet of Bug L55. The manually-written patch is shown in the \mycode{MANUAL PATCH} comment at Lines 6 and 7 while the patch by Nopol is shown in the \mycode{NOPOL PATCH} at Lines 8 and 9. The patch by Nopol is equivalent to the manually-written patch by developers.}
\label{fig:case-l55}
\end{figure}

Bug L55 in Commons Lang relates to a utility class for timing. The bug appears when the user stops a suspended timer: the stop time saved by the suspend action is overwritten by the stop action. Fig. \ref{fig:case-l55} presents the buggy method of Bug L55. In order to solve this problem, the assignment at Line 10 has to be done only if the timer state is running.

As shown in Fig. \ref{fig:case-l55}, the manually-written patch by the developer adds a precondition before the assignment at Line 10 and it checks that the current timer state is running (at Line 7). 
The patch by Nopol is different from the manually-written one. The Nopol patch compares the stop time variable to a integer constant (at Line 9), which is pre-defined in the program class and equals to $1$. In fact, when the timer is running, the stop time variable is equals to $-1$; when it is suspended, the stop time variable contains the stop time in millisecond. Consequently, both preconditions by developers and by Nopol are equivalent and correct. Despite being equivalent, the manual patch remains more understandable.
This bug is neither repaired by \jgenprog nor \jkali. To our knowledge, Nopol is the only approach that contains a strategy of adding preconditions to original statements, which does not exist in \jgenprog or \jkali. 
\textbf{To sum up, Bug L55 shows an example of a repaired bug, 1) that is in a hard-to-repair project (only Nopol succeeds) and 2) whose specification by the test suite is good enough to drive the synthesis of a correct patch.}

\textbf{Summary}. In this section, we have presented detailed case studies of three patches that are automatically generated  for three real-world bugs of Defects4J. Our case studies show that automatic repair approaches are able to repair real bugs. However, different factors, in particular the weakness of some test cases, yield clearly incorrect patches.

\section{Discussion}
\label{sect:discussion} 

\subsection{Threats to Validity}
\label{sec:threats}

\subsubsection{Benchmark}
Our results are obtained on four subject programs, from a benchmark in which there was no attempt to provide a representative set of fault classes.
Future experiments on other benchmarks are required mitigate this threat to the external validity.

We have taken the bugs of the Defects4J benchmark as they are and we did not modify the test cases. This is very important to decrease the following threats.
First, we measure the effectiveness of the repair tool as it would have be at the time when the bug was reported. We do not use other tests of information ``from the future''. 
It would introduce a potential experimental bias if we modify the benchmark, which has been set up independently of our experiment.

\subsubsection{Implementations of GenProg and Kali}
\label{threat:reimplementation}
\jgenprog and \jkali are the re-implementations of the GenProg and Kali algorithms in Java. There exists a threat that the implementations do not produce results that are as good as the original systems would (there is also a risk that the re-implementation produces better results). 
The authors of \jgenprog and \jkali carefully and faithfully reimplemented the original systems.  
Note this threat is absolute. Either one takes the risk or one give up any comparative evaluations between programming languages.
To find potential re-implementation issues, the systems are publicly available on Github for peer-review.

\subsubsection{Bias of assessing the correctness, readability, and difficulty}
In our work, each patch in Table \ref{tab:bug-correct} is validated by an analyst, which is one of the authors. An analyst manually identifies the correctness of a patch and labels the related readability and difficulty. Then, he discusses and validates the result of the patch analysis with another author, called a reviewer. 
Since patch correctness analysis on an unknown code base is a new kind of qualitative analysis, only explored by \cite{qi2015efficient}, we do not know its inherent problems and difficulty. 
This is why the methodological setup is a pilot one, where no independent validation has been attempted to measure inter-annotator agreement.
This results in a threat to the internal validity of the manual assessment.

We share our results online on the experiment Github repository to let readers have a clear idea of the difficulty of our analysis work and the correctness of generated patches (see Section \ref{subsect:answer-rq1}). For assessing the equivalence, one solution would be to use automatic technique, as done in mutation testing. However, whether the current equivalence detection techniques scale on large real Java code is an open question.

\subsubsection{Random nature of \jgenprog}
\label{threat:random}
\jgenprog, as the original GenProg implementation, has a random aspect: statements and mutations are randomly chosen during the search. Consequently, it may happen that a new run of \jgenprog yields different results.
There are two different threats. Our experiment may have over-estimated the repair effectiveness of GenProg (too many patches found) or underestimated the repair effectiveness (too few).

To analyze the overestimation threat, for each of the \nbFixedBugsByGenprog repaired bug, we have run jGenProg ten times with different seeds.
For 18/27 cases, a test-adequate patch was found in all 10 runs, in 6 cases, a test-adequate patch was found in at least 7 runs. In the remaining 3 cases, a patch was found in a minority of runs, and we can consider that our reference run was lucky to find at least one test-adequate patch. Note that this experiment is quite fast because the runs are mostly successful and hence terminate fast, much before the 3-hour time budget limit.

Analyze the underestimation threat is much more challenging, because if a bug is not repaired, it means that the trial lasted at least 3 hours (the time budget used in this experiment). Consequently, if the bug is really unrepairable, trying with $n$ different seeds would require $n \times 3$ hours of computation. For $224-27=197$ unrepaired bugs, this sets up a worst case bound of $197 \times 10 \times 3=5910$ hours of computation, i.e. 224 days, which is not reasonable, even split into parallel tasks.
Consequently, we decided to sample 5 unrepaired bugs and run with 10 different seeds (representing an experimental budget of $5 \times 10 \times 3=150$ hours).
The result was clear cut: no additional patches could be found. This suggests that the ``unrepairedness'' comes from the essential complexity of the problem and not from the accidental randomness of \jgenprog.

\subsubsection{Presence of multiple patches} 
In this experiment, we stop the execution of a repair system once the first patch is found. This is the patch that is manually analyzed. 
However, as also experienced by others, there are often several if not dozens of different patches per bug. 
It might happen that a correct patch lies somewhere in this set of generated patches. We did not manually analyze all generated patches because it would require months of manual work. This finding shows that there is a need for research on approaches that order the generated patches so as to reveal the most likely to be correct, this has been preliminarily explored by \cite{prophet}.

\subsection{Impact of Flaky Tests on Repair}
\label{sec:falky}

Our experiment uncovered one flaky test in Defects4J (C8).
We realized that flaky tests have a direct impact on automatic repair.
If the failing test case is flaky, the repair system might conclude that a correct patch has been found while it is actually correct.
If one of the passing test cases is flaky, the repair system might conclude that a patch has introduced a regression while it is not the case, this results in an underestimation of the effectiveness of the repair technique.

\subsection{Reflections on GenProg}

The largest evaluations of GenProg are by \cite{le2012systematic} and \cite{qi2015efficient}.
The former reports that 55/105 bugs are repaired (under the definition that the patch passes the test suite), while the latter argued that only 2/105 bugs are correctly repaired (under the definition that the patch passes the test suite and that the patch is correct and acceptable from the viewpoint of the developer). The difference is due to an experimental threat, the presence of under-specified bugs and a potential subjectiveness in the assessment.

In this paper, we find that our re-implementation of GenProg, \jgenprog correctly repairs \nbAnalyzedGenprogCorrect /\nbAllBugs (2.2\%) bugs. In addition, it uniquely repairs 4 bugs, such as M70 discussed in Section \ref{subsect:case-m70}.
We think that the difference in repair rate is probably due to the inclusion criteria of both benchmarks (GenProg and Defects4J). To our opinion, none of them reflect the actual distribution of all bug kinds and difficulty in the field.

However, the key finding is that having correctly and uniquely repaired bugs indicates that the core intuition of GenProg is valid, and that GenProg has to be a component of an integrated repair tool that would mix different repair techniques.

\subsection{On the Choice of Repair Techniques}

Given a new bug, can one choose the most appropriate repair technique? In practice, when a new bug is reported, one does not know in advance the type of its root cause. The root cause may be an incorrect condition, a missing statement, or something else. It means that given a failure (or a failing test case) there is no reason to choose one or another repair technique (say \jgenprog or Nopol). The pragmatic choice is to run all available techniques, whether sequentially or in parallel. However, there may be some information in the failure itself about the type of the root cause. Future results on the automatic identification of the type of root cause given a failure would be useful for repair, in order to guide the choice of repair technique.

\section{Related Work}
\label{sect:related}

\subsection{Real-World Datasets of Bugs}
\label{subsect:related-dataset}

The academic community has already developed many methods for finding and analyzing bugs. Researchers employ real-world bug data to evaluate their methods and to analyze their performance in practice. For instance, \cite{do2005supporting} propose a controlled experimentation platform for testing techniques. Their dataset is included in SIR database, which provides a widely-used testbed in debugging and test suite optimization.

\cite{dallmeier2007extraction} propose iBugs, a benchmark for bug localization obtained by extracting historical bug data.
BugBench by \cite{lu2005bugbench} and BegBunch by \cite{cifuentes2009begbunch} are two benchmarks that have been built to evaluate bug detection tools. 
The PROMISE repository (\cite{promise2015}) is a collection of datasets in various fields of software engineering. \cite{LeGoues15tse} have designed a benchmark of C bugs which is an extension of the GenProg benchmark.

In this experience report, we employ Defects4J by \cite{JustJE2014} to evaluate software repair. This database includes well-organized programs, bugs, and their test suites. The bug data in Defects4J has been extracted from the recent history of five widely-used Java projects. 
To us, Defects4J is the best dataset of real world Java bugs, both in terms of size and quality.
To our knowledge, our experiment is the first that evaluates automatic repair techniques via Defects4J.

\subsection{Test-Suite Based Repair Approaches}
\label{subsect:related-repair}

The idea of applying evolutionary optimization to repair derives from  \cite{DBLP:conf/cec/ArcuriY08}. Their work applies co-evolutionary computation to automatically generate bug patches. GenProg by \cite{le2012genprog} applies genetic programming to the AST of a buggy program and generates patches by adding, deleting, or replacing AST nodes. PAR by \cite{Kim2013} leverages patch patterns learned from human-written patches to find readable patches. RSRepair by \cite{qi2014strength} uses random search instead of genetic programming. This work shows that random search is more efficient in finding patches than genetic programming. Their follow-up work (\cite{DBLP:conf/icsm/QiML13}) uses test case prioritization  to reduce the cost of patch generation.

\cite{DBLP:conf/icst/DebroyW10} propose a mutation-based repair method inspired from mutation testing. This work combines fault localization with program mutation to exhaustively explore a space of possible patches. Kali by  \cite{qi2015efficient} has recently been proposed to examine the repair power of simple actions, such as statement removal. 

SemFix by \cite{nguyen2013semfix} is a notable constraint based repair approach. This approach provides patches for assignments and conditions by combining symbolic execution and code synthesis. Nopol by  \cite{demarco2014automatic} is also a constraint based method, which focuses on repairing bugs in \ourif conditions and missing preconditions. DirectFix by  \cite{mechtaev2015directfix} achieves the simplicity of patch generation with a Maximum Satisfiability (MaxSAT) solver to find the most concise patches. Relifix (\cite{relifix}) focuses on regression bugs. SPR (\cite{Long15}) defines a set of staged repair operators so as to early discard  many candidate repairs that cannot pass the supplied test suite and eventually to exhaustively explore a small and valuable search space.

Besides test-suite based repair, other repair setups have been proposed.  \cite{DBLP:journals/tse/0001FNWMZ14}  
proposed a contract based method for automatic repair. Other related repair methods include atomicity-violation fixing (e.g. \cite{DBLP:conf/pldi/JinSZLL11}), runtime error repair (e.g. \cite{DBLP:conf/pldi/LongSR14}), and domain-specific repair (e.g. \cite{DBLP:conf/icse/SamirniSAMTH12,gopinath2014data}).

\subsection{Empirical Investigation of Automatic Repair}
\label{subsect:related-analysis}

Beyond proposing new repair techniques, there is a thread of research on empirically investigating the foundations, impact and applicability of automatic repair. 

On the goodness of synthesized patches, 
\cite{Fry2012} conducted a study of machine-generated patches based on 150 participants and 32 real-world defects. Their work shows that machine-generated patches are slightly less maintainable than human-written ones. \cite{TaoKKX14} performed a similar study to study whether machine-generated patches assist human debugging.  \cite{monperrus2014critical} discussed in depth the acceptability criteria of synthesized patches. 

\cite{DBLP:journals/ese/MartinezM15} studied thousands of commits to mine repair models from manually-written patches. They later investigated (\cite{DBLP:conf/icse/MartinezWM14}) the redundancy assumption in automatic repair (whether you can repair bugs by rearranging existing code). 
\cite{zhong2015an} conducted a case study on over 9,000 real-world patches and found two important facts for automatic repair: for instance, their analysis outlines that some bugs are repaired with changing the configuration files.

A recent study by \cite{kong2015experience} compares four repair systems: including GenProg (\cite{le2012genprog}), RSRepair (\cite{qi2014strength}), Brute-force-based repair (\cite{le2012systematic}), and AE (\cite{weimer2013leveraging}). This work reported repair results on 119 seeded bugs and 34 real bugs from the Siemens benchmark and in the SIR repository. Our experiment is on a bigger set of real bugs coming from larger and more complex software applications.

\cite{Long2016analysis} have performed an analysis of the search space of their two systems SPR and Prophet on the benchmark of C bugs. They show that not all repair operators are equal with respect to finding the correct patch first. This is clearly different from what we do in this paper: 1) we consider a different benchmark (Defects4J), in a different programming language (Java), while they consider the Manybugs benchmark of C bugs.
2) they do not perform any manual analysis of their results.

\section{Conclusion}
\label{sect:conclusion} 

We have presented a large scale evaluation of automatic repair on a benchmark of real bugs in Java programs. Our experiment was conducted with three automatic repair systems on \nbAllBugs bugs in the Defects4J dataset. We find out that the systems under consideration can synthesize a patch for \nbFixedBugs out of \nbAllBugs. 
\emph{Since the dataset only contains real bugs from large-scale Java software, this is a piece of evidence about the applicability of automatic repair in practice.}

However, our manual analysis of all generated patches show that most of them are incorrect (beyond passing the test suite), because the overfit on the test data. Our experiment shows that overfitting also happens for Java software tested with a state-of-the-art testing infrastructure (as opposed to old-school C programs in \cite{qi2015efficient}); and also happens with Nopol, a synthesis based technique (and not only for Genprog and similar techniques).

Our results indicate two grand challenges for this research field: 1) producing systems that do not overfit (RQ2) and 2) producing systems that reason on the expected behavior beyond what is directly encoded in the test suite (RQ3).
There is also a third challenge, although not the focus on this experiment: during the exploratory phase, we have observed the presence of multiple patches for the same bug. This indicates a need for research on approaches that order the generated patches so as to reveal the most likely to be correct. Preliminary work is being done on this topic by \cite{prophet}.

Lastly, there is also a need for research on test suites. 
For instance, any approach that automatically enriches test suites with stronger assertions would have direct impact on repair, by preventing the synthesis of incorrect patches.

\newpage
\bibliographystyle{spbasic}  
\bibliography{references}

\end{document}